\documentclass{article}

\usepackage{arxiv}

\usepackage[utf8]{inputenc} 
\usepackage[T1]{fontenc}    
\usepackage{lmodern}        
\usepackage{hyperref}       
\usepackage{url}            
\usepackage{booktabs}       
\usepackage{amsfonts}       
\usepackage{nicefrac}       
\usepackage{microtype}      
\usepackage{graphicx}

\title{Analysis and sample size calculation within the responder
stratified exponential survival model}

\author{
    Samuel Kilian
   \\
    Institute of Medical Biometry \\
    Heidelberg University \\
  Heidelberg, Germany \\
  \texttt{\href{mailto:kilian@imbi.uni-heidelberg.de}{\nolinkurl{kilian@imbi.uni-heidelberg.de}}} \\
   \And
    Johannes Krisam
   \\
    Institute of Medical Biometry \\
    Heidelberg University \\
  Heidelberg, Germany \\
  \texttt{} \\
   \And
    Meinhard Kieser
   \\
    Institute of Medical Biometry \\
    Heidelberg University \\
  Heidelberg, Germany \\
  \texttt{} \\
  }

\providecommand{\tightlist}{%
  \setlength{\itemsep}{0pt}\setlength{\parskip}{0pt}}

\newlength{\cslhangindent}
\newlength{\csllabelwidth}
\newlength{\cslentryspacingunit} 
  {}%
  {\par}
\newenvironment{CSLReferences}[2] 
 {
  \setlength{\parindent}{0pt}
  \ifodd #1
  \let\oldpar\par
  \def\par{\hangindent=\cslhangindent\oldpar}
  \fi
  \setlength{\parskip}{#2\cslentryspacingunit}
 }%
 {}
\usepackage{calc}

\usepackage{amsmath}
\usepackage{booktabs}
\usepackage{longtable}
\usepackage{array}
\usepackage{multirow}
\usepackage{wrapfig}
\usepackage{float}
\usepackage{colortbl}
\usepackage{pdflscape}
\usepackage{tabu}
\usepackage{threeparttable}
\usepackage{threeparttablex}
\usepackage[normalem]{ulem}
\usepackage{makecell}
\usepackage{xcolor}
\begin{document}
\maketitle

\begin{abstract}
The primary endpoint in oncology is usually overall survival, where
differences between therapies may only be observable after many years.
To avoid withholding of a promising therapy, preliminary approval based
on a surrogate endpoint is possible. The approval can be confirmed later
by assessing overall survival within the same study. In these trials,
the correlation between surrogate endpoint and overall survival has to
be taken into account for sample size calculation and analysis. For a
binary surrogate endpoint, this relation can be modeled by means of the
responder stratified exponential survival (RSES) model proposed by Xia,
Cui, and Yang (2014). We derive properties of the model and confidence
intervals based on Maximum Likelihood estimators. Furthermore, we
present an approximate and an exact test for survival difference. Type I
error rate, power, and required sample size for both newly developed
tests are determined exactly. These characteristics are compared to
those of the logrank test. We show that the exact test performs best.
The power of the logrank test is considerably lower in some situations.
We conclude that the logrank test should not be used within the RSES
model. The proposed method for sample size calculation works well. The
interpretability of our proposed methods is discussed.
\end{abstract}


\newcommand{\Exp}{\mathrm{Exp}}
\newcommand{\Var}{\mathrm{Var}}
\newcommand{\Cov}{\mathrm{Cov}}
\newcommand{\asim}{\overset{\mathrm{appr}}{\sim}}
\newcommand{\lam}[2]{\lambda_{#1, #2}}
\newcommand{\thet}[2]{\theta_{#1, #2}}
\newcommand{\hthet}[2]{\hat\theta_{#1, #2}}
\newcommand{\pr}{^{\prime}}
\newcommand{\Rej}{\mathrm{Rej.}}
\newcommand{\CP}{\mathrm{CP}}
\newcommand{\CI}{\mathrm{CI}}

\hypertarget{introduction}{%
\section{Introduction}\label{introduction}}

Endpoints of clinical trials should be appropriate for answering the
research question, objectively measurable, and relevant for patients.
Thus, for proving efficacy of a new oncological therapy, the primary
endpoint is usually overall survival. The new therapy is compared to the
present gold standard. However, as therapies get better and diagnoses
are made in earlier stages, differences between therapies may only be
observable after many years. This may considerably delay approval of new
treatments and their application in practice.

To avoid withholding of a promising therapy, the Food and Drug
Administration (FDA) provides four different programs for expedited
development of new therapies (Wallach, Ross, and Naci 2018). One of them
is the \emph{Accelerated Approval} pathway, where the approval is based
on a surrogate endpoint. The approval is preliminary and has to be
confirmed later when the main endpoint can be assessed. Between 1992 and
2021, 278 preliminary accelerated approvals were granted by the FDA
after a median processing time of 6 months. Of these, 50\% were
confirmed later, 10\% were withdrawn, and 40\% are still ongoing (Food
and Drug Administration 2022).

In 2014, the FDA published a more detailed guidance (which was updated
in 2020) regarding the use of pathological complete response (pCR) as a
surrogate endpoint when approving a novel neoadjuvant treatment of
high-risk early-stage breast cancer (Food and Drug Administration 2020).
Although the appropriateness of pCR as surrogate endpoint is disputed
(Conforti et al. 2021), the guidance illustrates the use of a binary
surrogate endpoint for survival. When accelerated and final approval are
sought within the same study, the correlation between surrogate endpoint
and survival has to be taken into account for sample size calculation
and analysis. The relationship can be modeled by means of a conditional
survival proposed by Xia, Cui, and Yang (2014). They investigated the
correlation and assessed the power of the logrank test. Howevver, they
did not present methods for statistical testing, parameter estimation,
and sample size calculation. This gap constitutes a major hurdle for the
application of this approach.

The aim of our work is to develop methods for analysis and sample size
calculation within the conditional survival model by Xia, Cui, and Yang
(2014). In this article, we focus on the analysis of the survival
endpoint that is assessed at the end of the trial. Due to the
non-proportionality of the hazards within the model, standard methods
like the logrank test may not be the most powerful. The analysis of the
surrogate endpoint is built into the proposed testing strategy with
global Type I error control. However, the proposed sample size
calculation method is tailored to the final survival endpoint and does
not consider possible interim decisions after assessment of the
surrogate endpoint.

In Section 2, we define the model and explain the possible
constellations of marginal survival functions in a two-group model. We
derive Maximum Likelihood estimators and approximate confidence
intervals for the parameters in the third section. Based on that, an
approximate and an exact hypothesis test are derived in Section 4. Type
I error and power of these tests are evaluated in the fifth section. An
approximate sample size calculation method is derived and evaluated in
the sixth section. Furthermore, exact sample size calculation is
outlined. Finally, approximate sample sizes and exact power values are
calculated and compared for parameter values derived from clinical trial
data. The manuscript concludes with a discussion, while technical
details are given in the appendices.

\hypertarget{model}{%
\section{Statistical model}\label{model}}

\begin{figure}
\includegraphics[width=4.35in,]{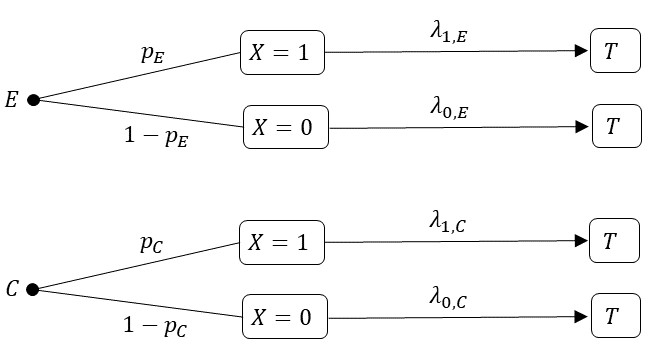} \caption{Two-group RSES model. Experimental (E) and control (C) group are each defined by a set of three parameters.}\label{fig:model}
\end{figure}

The \emph{responder stratified exponential survival} (RSES) model was
proposed by Xia, Cui, and Yang (2014). They did not specifically name
their model back then, we, however, decided to denote it as the RSES
model throughout the course of this manuscript. It models the survival
time of a patient as an exponentially distributed random variable with
the parameter depending on the response status of the patient. Formally,
the random variable \(X\) distinguishes the responders (\(X=1\)) from
the non-responders (\(X=0\)) and is Bernoulli distributed with
probability \(p\). The survival time \(T|X=1\) of a responder follows a
\(\mathrm{Exp}(\lambda_1)\)-distribution and the survival time \(T|X=0\)
of a non-responder follows a \(\mathrm{Exp}(\lambda_0)\)-distribution.
Figure \ref{fig:model} visualizes the model in two treatment groups. The
common distribution function of the random vector \((X, T)\) is
\[F_{p, \lambda_1, \lambda_0}(x, t) = x \cdot p \cdot \left(1-\exp(-\lambda_1 t)\right) + (1-x) \cdot (1-p) \cdot \left(1-\exp(-\lambda_0 t)\right).\]
The common density function is
\[f_{p, \lambda_1, \lambda_0}(x, t) = x \cdot p \cdot \lambda_1\exp(-\lambda_1 t) + (1-x) \cdot (1-p) \cdot \lambda_0\exp(-\lambda_0 t).\]

By integrating over \(x\), we get the marginalized distribution of the
survival time:
\[\tilde F_{p, \lambda_1, \lambda_0}(t) = p \cdot \left(1-\exp(-\lambda_1 t)\right) + (1-p) \cdot \left(1-\exp(-\lambda_0 t)\right)\]

Thus, the survival function is

\begin{align*}
S_{p, \lambda_1, \lambda_0}(t) &= 1 - \tilde F_{p, \lambda_1, \lambda_0}(t)\\
&= p\cdot \exp(-\lambda_1 t) + (1-p)\cdot\exp(-\lambda_0 t).
\end{align*}

When comparing the experimental group \(E\) with the control group
\(C\), differences within the three-parameter model may not be easily
interpretable. Specifically, a difference in parameter sets does not
imply a global survival benefit of one group neither does it imply a
survival difference between groups.

Let \(p_i, \lambda_{1, i}, \lambda_{0, i}\) be the respective parameter
sets of the groups as shown in Figure \ref{fig:model}. For better
readability, \(S_{p_i, \lambda_{1, i}, \lambda_{0, i}}\) will be
abbreviated as \(S_i\). We can distinguish three cases of the relation
of \(S_E\) and \(S_C\):

\begin{enumerate}
\def\labelenumi{\arabic{enumi}.}
\tightlist
\item
  Completely equal: \(S_E(t) = S_C(t) \quad \forall t \ge 0\)
\item
  Uniformly different: \(S_E(t) \ne S_C(t) \quad \forall t > 0\)
\item
  Crossing: not completely equal but \(\exists\ t > 0\) such that
  \(S_E(t) = S_C(t)\)
\end{enumerate}

See Appendix \ref{surv-diff} for a detailed analysis of these cases.

\begin{figure}
\includegraphics{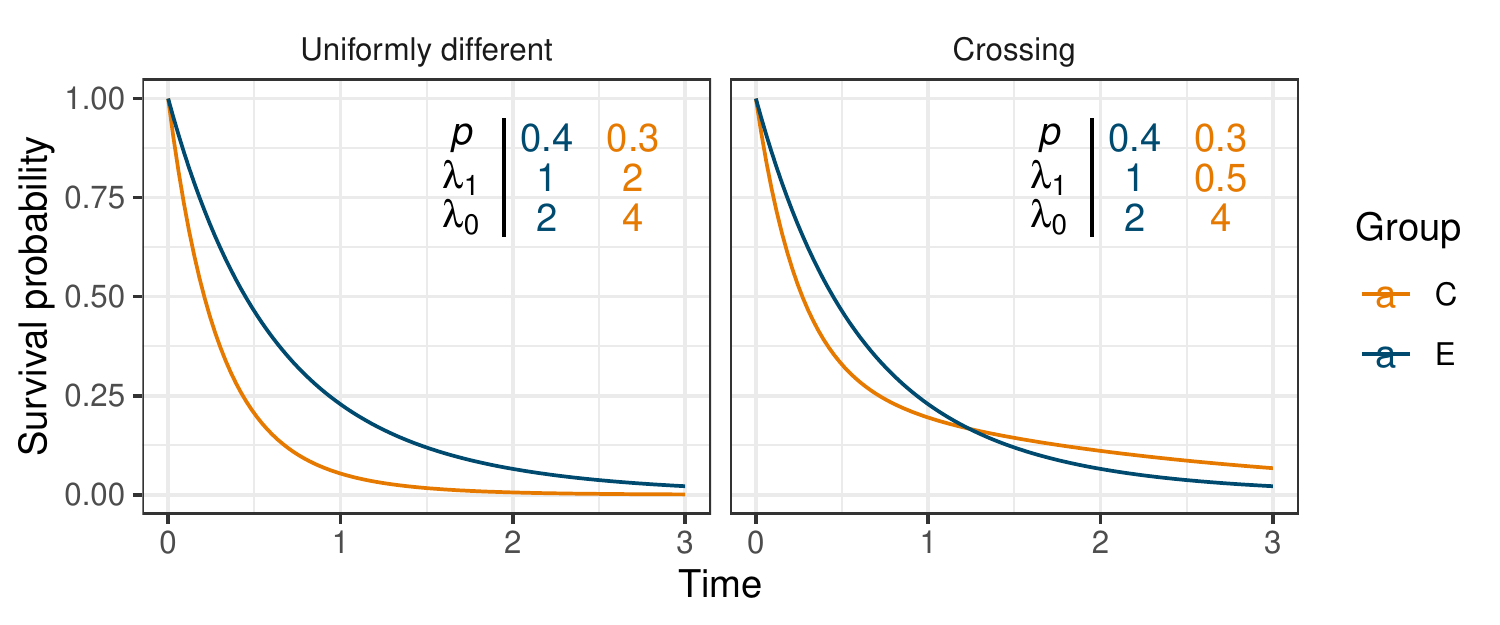} \caption{Survival functions of two different two-group models.}\label{fig:surv-diff}
\end{figure}

Figure \ref{fig:surv-diff} shows two examples. In the first, the
experimental group has a higher response probability and better survival
of each responders and non-responders. Thus, survival in the
experimental group is uniformly better. In the second example, responder
survival is better in the control group. This advantage comes into
effect after a certain time which leads to crossing survival curves.

\hypertarget{sec-mle}{%
\section{Maximum Likelihood estimation}\label{sec-mle}}

In this section, estimators of the model parameters are derived by the
Maximum Likelihood method. Suppose we observe \(n\) realisations
\((x_i, t_i)\) of response status and survival time. For simplicity, we
assume that there is no censoring. The inclusion of censoring will be
addressed in the Discussion. Let \(k = \sum_{i=1}^n x_i\) be the number
of responders. We arrange our observations such that
\(x_1 = \dots = x_{k} = 1\) and \(x_{k+1} = \dots = x_n = 0\). Denote
\(\theta_i = \log(\lambda_i)\). Then the Maximum Likelihood estimates of
the parameters \(p, \theta_1\) and \(\theta_0\) are

\begin{align*}
\hat p &= \frac{k}{n},\\
\hat \theta_1 &= -\log\left( \frac{1}{k}\sum\limits_{i = 1}^{k} t_i\right),\\
\hat \theta_0 &= -\log\left( \frac{1}{n-k}\sum\limits_{i = k+1}^{n} t_i\right).
\end{align*}

See Appendix \ref{mle-derivation} for a detailed derivation of the MLEs
and their asymptotic covariance structure. Note that the estimation of
\(\hat \theta_1\) is only well-defined if \(k \ne 0\). In the case
\(k = 0\) there is no unique Maximum Likelihood estimator for
\(\theta_1\). The same holds for \(\theta_0\) if \(k = n\).

The asymptotic distribution of the MLE is multivariate normal with the
true parameter vector as mean and the inverse Fisher information matrix
as covariance matrix.\\
This yields \begin{align*}
\mathrm{Var}(\hat p) &\approx \frac{p\cdot(1-p)}{n},\\ 
\mathrm{Var}(\hat \theta_1) &\approx \frac{1}{np},\\
\mathrm{Var}(\hat \theta_0) &\approx \frac{1}{n(1-p)},\\
\mathrm{Cov}(\hat p, \hat \theta_i) &\approx 0,\\
\mbox{and } \mathrm{Cov}(\hat \theta_1, \hat\theta_0) &\approx 0.
\end{align*} In particular, the MLEs of the different parameters are
asymptotically uncorrelated.

Also, the exact distribution of the MLEs can be given explicitly:
\(n \cdot \hat p = k\) follows a binomial distribution with parameters
\(n\) and \(p\). Conditional on \(k\), \(\sum\limits_{i = 1}^{k} t_i\)
follows a \(\Gamma(k, \lambda_1)\)-distribution and thus
\[\exp(-\hat\theta_1) \sim \Gamma(k, k\lambda_1).\]

\begin{figure}
\includegraphics{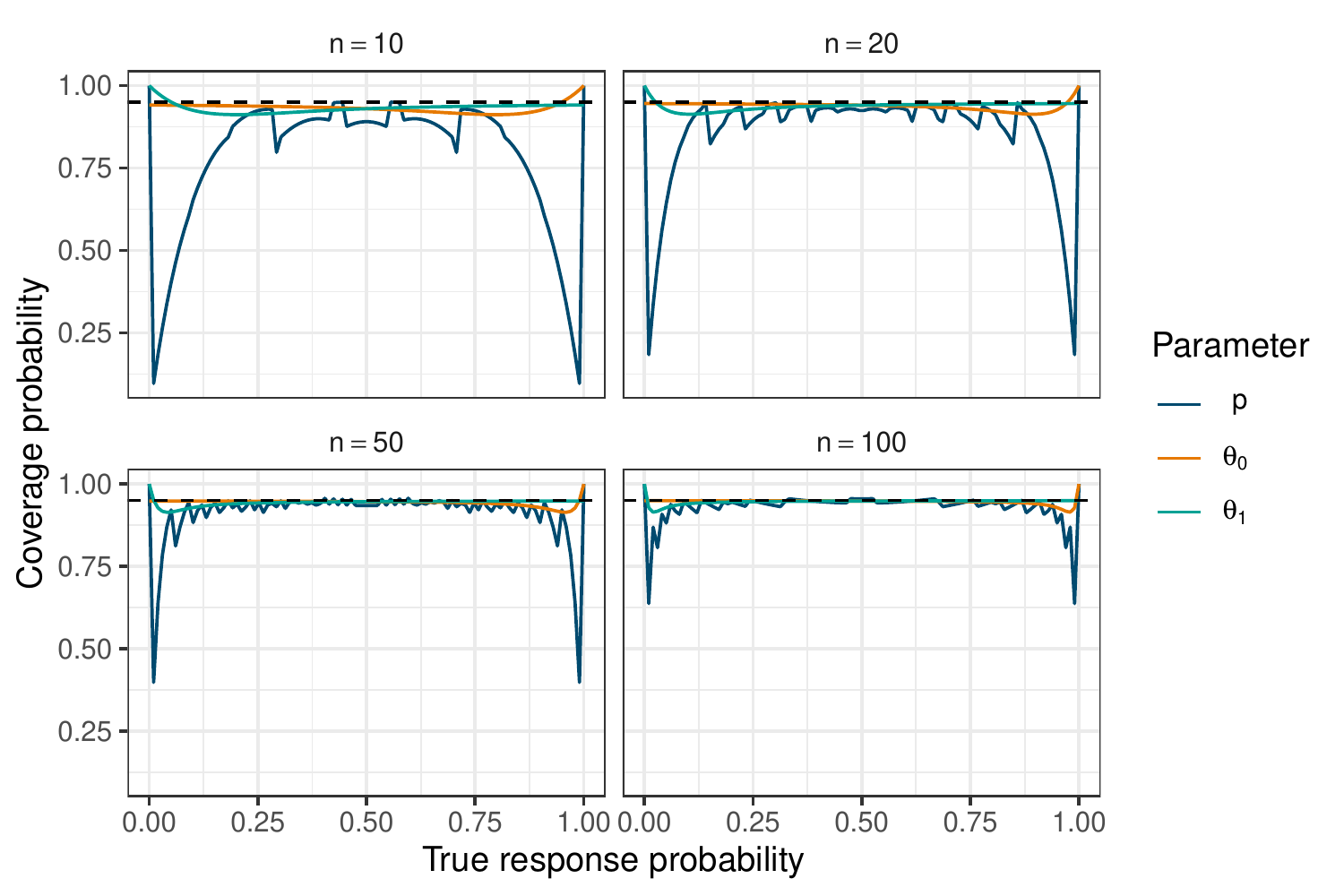} \caption{Exact coverage probability of approximate confidence intervals for various values of $p$ and sample sizes.}\label{fig:ci-cov-prob-fig}
\end{figure}

Asymptotic confidence intervals for the parameters can be constructed by
using the asymptotic normality of the MLEs. For \(p\) this yields the
well known normal approximation confidence interval
\[\mathrm{CI}_p(\hat p) = \hat p \pm z_{1-\alpha/2} \cdot \sqrt{\hat p(1-\hat p)/n}.\]
For \(\theta_1\) we obtain
\[\mathrm{CI}_{\theta_1}(\hat\theta_1, \hat p) = \hat\theta_1 \pm z_{1-\alpha/2} \cdot \sqrt{1/(n \hat p)},\]
and for \(\theta_0\)
\[\mathrm{CI}_{\theta_0}(\hat\theta_0, \hat p) = \hat\theta_0 \pm z_{1-\alpha/2} \cdot \sqrt{1/\left(n (1 - \hat p)\right)}.\]
If \(\hat p = 0\) (or \(\hat p = 1\)), the confidence interval for
\(\theta_1\) (or \(\theta_0\)) cannot be computed and is formally set to
\((-\infty, \infty)\) to enable calculation of coverage probabilities.
See Appendix \ref{ci-cov-prob} for details of the calculation. Figure
\ref{fig:ci-cov-prob-fig} shows the coverage probability of the
asymptotic 95\% confidence intervals for different sample sizes and true
response probabilites. The coverage probability of
\(\mathrm{CI}_{\theta_1}\) and \(\mathrm{CI}_{\theta_0}\) is pretty good
overall. The coverage probability of \(\mathrm{CI}_p\) is generally
lower and much too low for extreme probabilites and small sample sizes.

\hypertarget{hypothesis-testing}{%
\section{Hypothesis testing}\label{hypothesis-testing}}

There are different ways to formulate a null hypothesis regarding the
RSES model. In this article, we are testing the difference of parameter
sets. We consider the comparison of an experimental group \(E\) and a
control group \(C\) with respective parameter triples
\((p_E, \theta_{1, E}, \theta_{0, E})\) and
\((p_C, \theta_{1, C}, \theta_{0, C})\). We want to test the null
hypothesis
\[H_0: p_C = p_E, \theta_{1, C} = \theta_{1, E} \mbox{ and } \theta_{0, C} = \theta_{0, E}.\]

Note that this is not a neccessary condition for equality of the
marginal survival distributions as we saw in Section \ref{model}. This
procedure aims at detecting any group difference within the parametric
model. In particular, it is not meant to make inference about
differences of single parameters because they can only be interpreted
and translated to survival difference when considered as a triple.

Our global null hypothesis \(H_0\) is an intersection of three local
null hypotheses: \begin{align*}
H_{p, 0}\mbox{: }&p_C = p_E\\
H_{\theta_1, 0}\mbox{: }&\theta_{1, C} = \theta_{1, E}\\
H_{\theta_0, 0}\mbox{: }&\theta_{0, C} = \theta_{0, E}
\end{align*}

Firstly, we will construct asymptotic Wald-like tests for the local test
problems. By splitting the desired significance level \(\alpha\) into
local levels \(\tilde\alpha\), this will give an asymptotic test for the
global null hypothesis. Secondly, we will construct an exact test based
on the exact distribution of the test statistics.

\hypertarget{approximate-test}{%
\subsection{Approximate test}\label{approximate-test}}

We construct test statistics for the local hypotheses by standardizing
the difference between the MLEs of both groups. It is divided by the
standard deviation of the difference. Here, the unknown true response
probabilities are replaced by their MLEs under the null hypothesis. The
MLE of the response probability under \(H_0\) is
\(\hat p = \frac{n_E \hat p_E + n_C \hat p_C}{n_E + n_C}\).

For the test of \(H_{p, 0}\), this yields the well-known two-sample
binomial \(z\)-test with test statistic
\[T_p = \frac{\hat p_E - \hat p_C}{\sqrt{\hat p(1- \hat p)(\frac{1}{n_E} + \frac{1}{n_C})}}.\]
If \(\hat p \in \{0, 1\}\) (which means that everyone or no one is a
responder), we set \(T_p = 0\) because we would not reject \(H_{p, 0}\)
in this case.

The test statistic for the local test of \(H_{\theta_1, 0}\) is
\[T_{\theta_1} = \frac{\hat\theta_{1, E} - \hat\theta_{1, C}}{\sqrt{\frac{1}{\hat p}(\frac{1}{n_E} + \frac{1}{n_C})}}.\]
If the number of responders \(k_E, k_C\) in one of the groups equals
zero, we set \(T_{\theta_1} = 0\) because we cannot make a statement
about survival of responders in one of the groups and thus would not
reject \(H_{\theta_1, 0}\).

Analogously, the test statistic for the local test of
\(H_{\theta_0, 0}\) is
\[T_{\theta_0} = \frac{\hat\theta_{0, E} - \hat\theta_{0, C}}{\sqrt{\frac{1}{(1-\hat p)}(\frac{1}{n_E} + \frac{1}{n_C})}}.\]
Again, if \(k_E = n_E\) or \(k_C = n_C\), we set \(T_{\theta_0} = 0\)
because we cannot make a statement about survival of non-responders in
one of the groups and thus would not reject \(H_{\theta_0, 0}\).

The three test statistics are asymptotically standard normally
distributed under their respective null hypothesis. Since the MLEs are
asymptotically uncorrelated, the three test statistics are also
asymptotically uncorrelated. Thus, we test the intersection of the three
local hypotheses by assuming independence of the test statistics and
testing every hypothesis at level
\(\tilde\alpha = 1-\sqrt[3]{1-\alpha}\). The global hypothesis \(H_0\)
can be rejected if at least one of the local hypotheses can be rejected.
This will asymptotically control the Type I error rate at the level
\(\alpha\). The level \(\tilde \alpha\) is based on an equal split of
\(\alpha\). However, the procedure can easily be adapted to other
allocation methods.

\hypertarget{sec-ex-test}{%
\subsection{Exact test}\label{sec-ex-test}}

To construct an exact test, we test \(H_{\theta_1, 0}\) and
\(H_{\theta_0, 0}\) conditionally on \(k_E\) and \(k_C\). By doing so,
the test statistic \(T_{\theta_1}\) becomes a monotone transformation of
the simplified test statistic
\[\tilde T_{\theta_1} := \frac{k_C\bar T_{1,C}}{k_E\bar T_{1, E}}.\]
\(\frac{\lambda_{1, C}}{\lambda_{1, E}}\tilde T_{\theta_1}\) follows a
beta prime distribution:
\[\frac{\lambda_{1, C}}{\lambda_{1, E}}\tilde T_{\theta_1} \sim \beta'(k_C, k_E).\]
This distribution is also known as beta distribution of the second kind
(Johnson, Kotz, and Balakrishnan 1995) and can be defined by its density
\[f_{\alpha, \beta}(x) = \frac{x^{\alpha-1}(1+x)^{-\alpha-\beta}}{B(\alpha,\beta)} \quad\quad\mbox{for }x > 0,\]
where \(B\) is the Beta function. Thus, under the null hypothesis
\(\lambda_{1, E}=\lambda_{1, C}\), the exact distribution of
\(\tilde T_{\theta_1}\) is independent of
\(\lambda_{1, E}, \lambda_{1, C}\). The same can be done for
\(\tilde T_{\theta_0}\).

\(H_{p, 0}\) can be tested exactly by using any exact test for the
comparison of two binomial proportions, e.g.~Boschloo's test (Boschloo
1970). For consistency, we use an exact test based on the test statistic
\(T_p\), which Mehrotra, Chan, and Berger (2003) named \emph{Z-Pooled
Exact Unconditional Test}.

Hence, the exact test procedure consists in computing the exact local
p-value \(p_p\) for the test of \(H_{p, 0}\) and the exact local
p-values \(p_{\theta_1}\) and \(p_{\theta_0}\) conditionally on \(k_E\)
and \(k_C\). If one of the p-values is smaller than the local level
\(\tilde\alpha\), we reject the global hypothesis \(H_0\). This
procedure yields an exact test for \(H_0\) controlling the Type I error
rate at the level \(\alpha\) (see Appendix
\ref{exact-test-keeps-level}).

\hypertarget{assessment-of-test-characteristics-and-comparison-to-logrank-test}{%
\section{Assessment of test characteristics and comparison to logrank
test}\label{assessment-of-test-characteristics-and-comparison-to-logrank-test}}

In this section, we analyse Type I error rate and power of the
approximate and the exact test and compare it to the logrank test and
the stratified logrank test. See Appendix \ref{logrank-tests} for
details of the used logrank test statistics. Note that stratifying the
logrank test for response status deliberately ignores a survival benefit
originating from a response benefit. Thus, the stratified logrank test
is not appropriate in our setting. However, we included it for
comparison purposes because Xia, Cui, and Yang (2014) considered it as
well.

Rejection probabilities for the proposed tests were calculated exactly.
For the logrank tests, rejection probabilities were estimated by
simulation with \(10^5\) runs per scenario, resulting in a standard
error of 0.0007 for the Type I error and a maximal standard error of
0.0016 for the power.

\hypertarget{assessment-of-type-i-error-rate}{%
\subsection{Assessment of Type I error
rate}\label{assessment-of-type-i-error-rate}}

\begin{figure}
\includegraphics{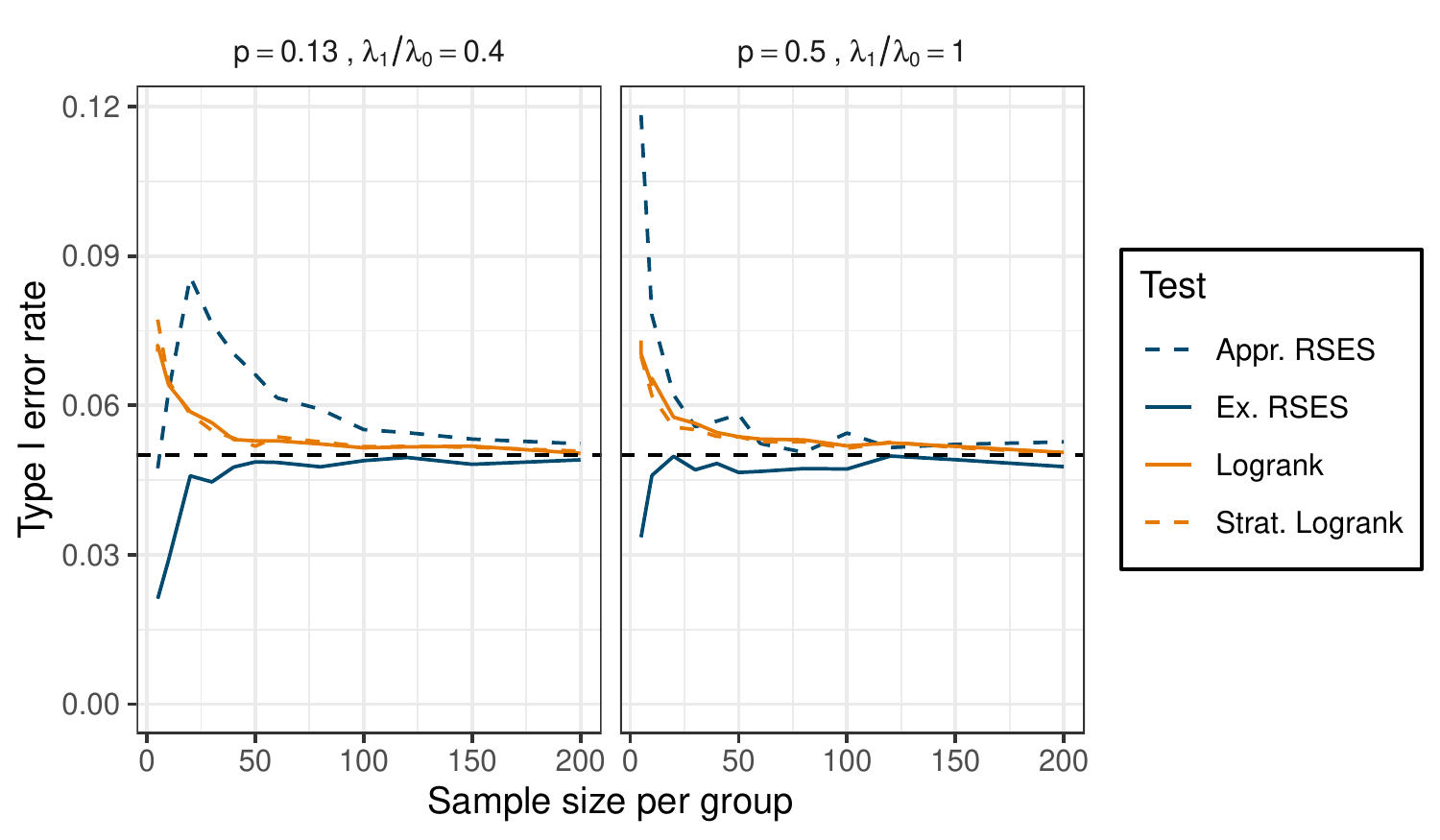} \caption{Type I error rate for various sample sizes in different scenarios.}\label{fig:t1e}
\end{figure}

We evaluated the Type I error rate in various scenarios and found
similar results in all of them. Thus, we chose two representative
scenarios for illustration. In the first setting, survival of responders
and non-responders is equal and the response probability equals 0.5 in
both groups. In the second scenario, survival of responders is better,
with a hazard ratio of 0.4, and the response probability equals 0.13 in
both groups. No difference with regard to survival is assumed between
experimental and control group. Figure \ref{fig:t1e} shows Type I error
rates for sample sizes per group from 5 to 200. The exact test always
adheres to the nominal level and exploits it even for small sample
sizes. The two logrank tests and the approximate RSES test exceed the
nominal level for small sample sizes but the two logrank tests converge
faster to the desired value. The approximate test performs worse for
smaller response probabilities. This is due to the fact that the normal
approximation of \(T_p\) is worse. Furthermore, the expected size of the
responder stratum is smaller and thus the normal approximation of
\(T_{\theta_1}\) is worse. The same holds true for response
probabilities near 1.

Interestingly, Type I error rate of the asymptotic test is lower for
very small sample sizes and response probabilities. This is because in
these cases it is likely that one of the groups does not include any
responders, which prevents rejection of \(H_{\theta_1, 0}\). The same
holds true for response probabilities near 1.

\hypertarget{sec-power}{%
\subsection{Assessment of Power}\label{sec-power}}

We consider five different scenarios for the evaluation of the power. In
all scenarios, responder survival is better than non-responder survival
with a hazard ratio of \(\lambda_1 / \lambda_0 = 0.4\). In the first two
scenarios (\emph{+resp}), survival benefit of the experimental group is
solely due to a higher response probability (\(p_E = 0.26, 0.52\)
vs.~\(p_C = 0.13\)). In Scenarios 3 and 4 (\emph{+resp +surv}), there is
additionally a survival benefit of the experimental group within the
strata. That means that responders in the experimental group have a
better survival than responders in the control group and non-responders
in the experimental group have a better survival than non-responders in
the control group. In Scenario 5 (\emph{+surv}), the response
probabilities are equal in both groups. The survival benefit of the
experimental group is solely due to a better survival of both responders
and non-responders.

\begin{figure}
\includegraphics{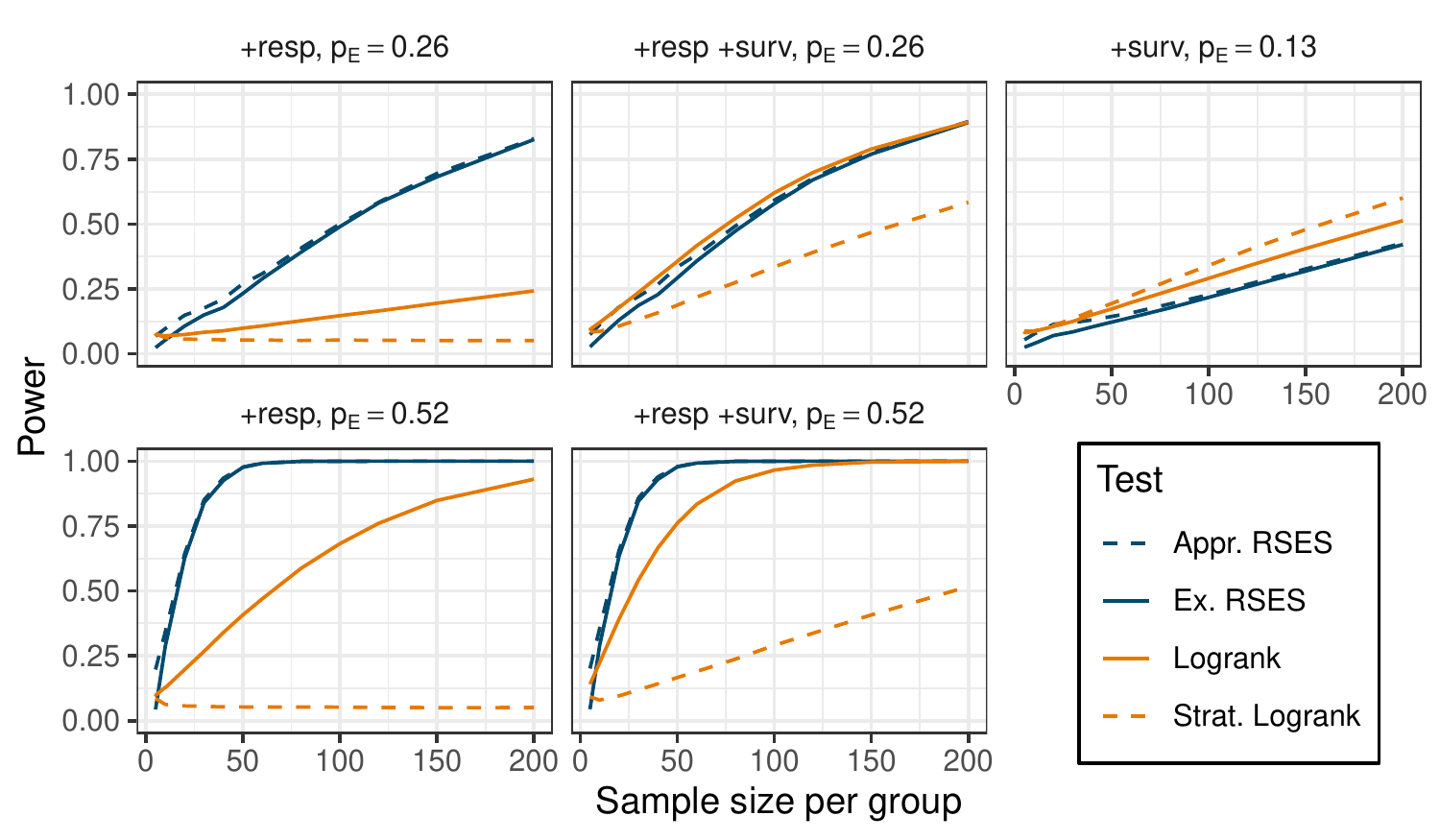} \caption{Power with respect to sample size in different treatment effect scenarios.}\label{fig:power}
\end{figure}

Figure \ref{fig:power} shows power for sample sizes per group from 5 to
200. In all scenarios, the approximate RSES test has slightly greater
power than the exact test. However, the difference is not meaningful.

When survival benefit is solely due to response benefit (\emph{+resp}),
the RSES tests are much more powerful than the logrank test. Since the
stratified logrank test only considers survival differences within the
response strata, its power equals the significance level.

The higher the survival benefit within the strata compared to the
response benefit, the better perform the logrank tests compared to the
RSES tests. This is because they don't spend significance level to test
response difference. When there is no response difference at all
(\emph{+surv}), the logrank tests are more powerful.

Interestingly, power of the stratified logrank test is not always
monotonically increasing for very small sample sizes. This is because in
these cases it is likely that one of the strata is empty. Then the
stratified logrank test becomes a usual logrank test restricted to the
non-empty stratum and can be more powerful than a stratified logrank
test on two slightly bigger strata.

\hypertarget{sec-ss-calc}{%
\section{Sample size calculation}\label{sec-ss-calc}}

Since both the approximate and the exact RSES test are based on the same
test statistics with the same asymptotic distributions, approximate
sample size calculation is valid for both tests. Exact sample size
calculation differs for both tests and is only presented for the exact
test.

\hypertarget{approximate-sample-size-calculation}{%
\subsection{Approximate sample size
calculation}\label{approximate-sample-size-calculation}}

Firstly, we derive for the three local tests approximate formulas for
the probabilities to falsely accept the respective null hypothesis. This
can be done using the asymptotic normality of the test statistics. Let
\(n_E, n_C\) be the sample sizes and
\(p_i^{\prime}, \theta_{j, i}^{\prime}\) the specified parameter values
under the assumed alternative hypothesis for \(i = E,C, j = 0,1\). Then
the approximate probabilities for falsely accepting the respective null
hypothesis of the three local tests are \begin{align*}
\beta_p &= \Phi\left( \frac{z_{1-\frac{\tilde\alpha}{2}} \cdot \sqrt{\tilde p(1- \tilde p)(\frac{1}{n_E} + \frac{1}{n_C})} - (p_E^{\prime}-p_C^{\prime})}{\sigma_p^{\prime}}\right),\\
\beta_{\theta_1} &= \Phi\left( \frac{z_{1-\tfrac{\alpha}{2}} \cdot \sqrt{\tfrac{1}{\tilde p}\cdot(\tfrac{1}{n_E} + \tfrac{1}{n_C})} - (\theta_{1, C}^{\prime}-\theta_{1, E}^{\prime})}{\sigma_{\theta_1}^{\prime}}\right),\\
\beta_{\theta_0} &= \Phi\left( \frac{z_{1-\tfrac{\alpha}{2}} \cdot \sqrt{\tfrac{1}{(1-\tilde p)}\cdot(\tfrac{1}{n_E} + \tfrac{1}{n_C})} - (\theta_{0, C}^{\prime}-\theta_{0, E}^{\prime})}{\sigma_{\theta_0}^{\prime}}\right),
\end{align*} with
\[\tilde p := \frac{n_E p_E^{\prime}+ n_C p_C^{\prime}}{n_E + n_C}.\]

See Appendix \ref{appr-acc-prob} for the derivation of the formulas. Due
to the asymptotic independence of the three test statistics, the
probability to not reject \(H_0\), i.e.~to accept all three local null
hypotheses simultaneously, is approximately equal to the product of the
acceptance probabilities of the three local null hypotheses.

Let \(r = n_E / n_C\) be the desired sample size ratio. We specify power
\(1-\beta\), significance level \(\alpha\), and all distribution
parameters, and set the local level to
\(\tilde\alpha = 1-\sqrt[3]{1-\alpha}\). Then, due to
\({n_E = r \cdot n_C}\), the acceptance probabilities of the three local
tests can be viewed as functions of \(n_C\). Thus, the required control
group sample size \(n_C\) is the solution of the equation
\[\beta_p(n_C) \cdot \beta_{\theta_1}(n_C) \cdot \beta_{\theta_0}(n_C) = \beta,\]
which can be determined numerically.

This approach also leaves the possibility of splitting Type I or Type II
error rate differently to weight certain hypotheses. The calculation
method can easily be adapted to such changes.

\begin{figure}
\includegraphics{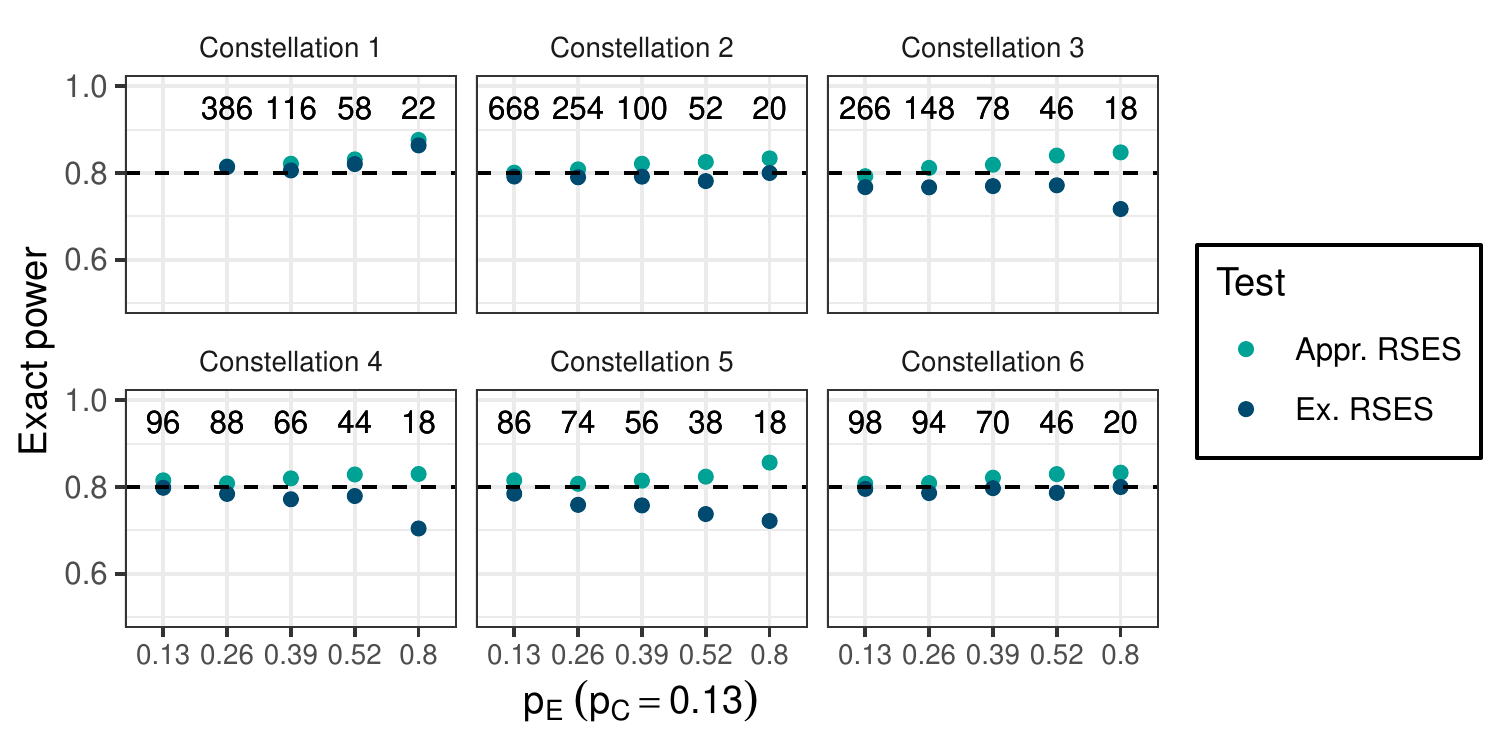} \caption{Exact power of approximate sample sizes in various scenarios. Calculated sample sizes are shown in the top row.}\label{fig:samplesize}
\end{figure}

Figure \ref{fig:samplesize} shows the exact power of exact and
approximate test using the approximate sample size calculation in 29
different scenarios. We set the non-responder hazard to
\({\gamma := 0.142}\) and the sample size ratio to \(r = 1\). We
consider the following six constellations of responder and non-responder
survival:

\begin{enumerate}
\def\labelenumi{\arabic{enumi}.}
\tightlist
\item
  Equal survival in all strata:
  \(\lambda_{0, E} = \lambda_{0, C} = \lambda_{1, E} = \lambda_{1, C} = \gamma\)
\item
  Better survival of responders in experimental group (1):
  \(\lambda_{0, C} = \lambda_{0, E} = \lambda_{1, C} = \gamma\) and
  \(\lambda_{1, E} = \gamma / 2\)
\item
  Better survival of responders in experimental group (2):
  \(\lambda_{0, C} = \lambda_{0, E} = \lambda_{1, C} = \gamma\) and
  \(\lambda_{1, E} = \gamma / 3\)
\item
  Better survival of responders and non-responders in experimental
  group: \(\lambda_{0, C} = \lambda_{1, C} = \gamma\) and
  \(\lambda_{0, E} = \lambda_{1, E} = \gamma / 2\)
\item
  Better survival of non-responders in experimental group, even better
  survival of responders in experimental group:
  \(\lambda_{0, C} = \lambda_{1, C} = \gamma, \lambda_{0, E} = \gamma/2\)
  and \(\lambda_{1, E} = \gamma / 3\)
\item
  Better survival of responders in control group, better survival of
  non-responders in experimental group, even better survival of
  responders in experimental group:
  \(\lambda_{0, C} = \gamma, \lambda_{1, C} = \gamma/2, \lambda_{0, E} = \gamma/2\)
  and \(\lambda_{1, E} = \gamma / 3\)
\end{enumerate}

In each of these constellations, response probability in the control
group is \(p_C = 0.13\). We consider five different response
probabilities in the experimental group:
\({p_E = 0.13,\ 0.26,\ 0.39,\ 0.52,\ 0.8}\). When response probabilities
are equal in the first constellation, there is no group difference.
Hence, no sample size can be calculated in this case.

We see that the approximate sample size calculation method works pretty
well for the approximate test and slightly underestimates the required
sample size for the exact test in most scenarios. This is consistent
with the small power advantage of the approximate test seen in Section
\ref{sec-power}. If the response probabilities and responder and
non-responder survival differ strongly between the two groups, the power
drops considerably below the desired value. This correlates with an
increased Type I error rate of the approximate test in such cases.

\hypertarget{exact-sample-size-calculation}{%
\subsection{Exact sample size
calculation}\label{exact-sample-size-calculation}}

Exact sample size calculation can be done iteratively because power can
be calculated exactly. The power of the exact RSES test of a specific
alternative \(H_A\) with parameters
\(p_{i, A}, \theta_{1, i,A}, \theta_{0, i,A}\) (\({i = E, C}\)) is given
by

\begin{align*}
P_{H_A}(\mathrm{Rej.}H_0) &= P_{p_{E, A}, p_{C,A}}(\mathrm{Rej.}H_{p, 0}) \\
& \quad + \sum_{(k_E, k_C) \notin R_p} 
f_{p_{E, A}, p_{C,A}}(k_E, k_C) \cdot \left(1 - u_1(k_E, k_C) \cdot u_0(n_E - k_E, n_C - k_C)\right).
\end{align*}

Here, \(R_p\) is the rejection region of the exact test of \(H_{p, 0}\).
Furthermore,
\(u_1(k_E, k_C) = u_{\lambda_{1, E,A}, \lambda_{1, C,A}}(k_E, k_C)\) is
the probability to not reject \(H_{\theta_1, 0}\) if the true parameters
are \(\lambda_{1, E,A}, \lambda_{1, C,A}\) and the group sizes are
\(k_E, k_C\). If \(k_E = 0\) or \(k_C = 0\), we have
\(u_{\lambda_{E, A}, \lambda_{C, A}}(k_E, k_C) = 1\) (``empty group
cases''). In all other cases we have

\begin{align*}
&u_{\lambda_{E, A}, \lambda_{C, A}}(k_E, k_C) \\
&= P_{\lambda_{E, A}, \lambda_{C, A}}\left(-c_\lambda(k_E, k_C, \tilde\alpha) \le \tilde T_{\theta_1} \le c_\lambda(k_E, k_C, \tilde\alpha)\right)\\
&= P\left(\frac{\lambda_{1, C} k_C}{\lambda_{1, E} k_E} \cdot \exp\left(-c_\lambda(k_E, k_C, \tilde\alpha)\right) \le \frac{\lambda_{1, C} k_C}{\lambda_{1, E} k_E} \cdot \exp(\tilde T_{\theta_1}) \le \frac{\lambda_{1, C} k_C}{\lambda_{1, E} k_E} \cdot \exp\left(c_\lambda(k_E, k_C, \tilde\alpha)\right)\right)\\
&= F_{\beta^{\prime}(k_C, k_E)}\left(\frac{\lambda_{1, C} k_C}{\lambda_{1, E} k_E} \cdot \exp\left(c_\lambda(k_E, k_C, \tilde\alpha)\right)\right) - F_{\beta^{\prime}(k_C, k_E)}\left(\frac{\lambda_{1, C} k_C}{\lambda_{1, E} k_E} \cdot \exp\left(-c_\lambda(k_E, k_C, \tilde\alpha)\right)\right).
\end{align*}

\(c_\lambda(k_E, k_C, \tilde\alpha)\) is the critical value of the
two-sided test of \(H_{\theta_1, 0}\) conditionally on the numbers of
responders \(k_E, k_C\). It is defined by the equation
\[1 - F_{\beta^{\prime}(k_C, k_E)}\left(k_C/k_E\cdot\exp\left( c_\lambda(k_E, k_C, \tilde\alpha)\right)\right)
+ F_{\beta^{\prime}(k_C, k_E)}\left(k_C/k_E\cdot\exp\left(- c_\lambda(k_E, k_C, \tilde\alpha)\right)\right)
= \tilde\alpha\] and can be calculated numerically.

Analogously,
\(u_0(n_E - k_E, n_C - k_C) = u_{\lambda_{0, E,A}, \lambda_{0, C,A}}(n_E - k_E, n_C - k_C)\)
is the probability to not reject \(H_{\theta_0, 0}\). Thus, the exact
sample size can be calculated iteratively as follows:

\begin{enumerate}
\def\labelenumi{\arabic{enumi}.}
\tightlist
\item
  Start with the approximate sample size.
\item
  Calculate exact power \(P_{H_A}(\mathrm{Rej.}H_0)\).
\item
  Increase sample size if power is too low, decrease sample size if
  power is too high.
\item
  Iterate steps 2 and 3.
\end{enumerate}

\hypertarget{example}{%
\section{Example}\label{example}}

\begin{figure}
\includegraphics{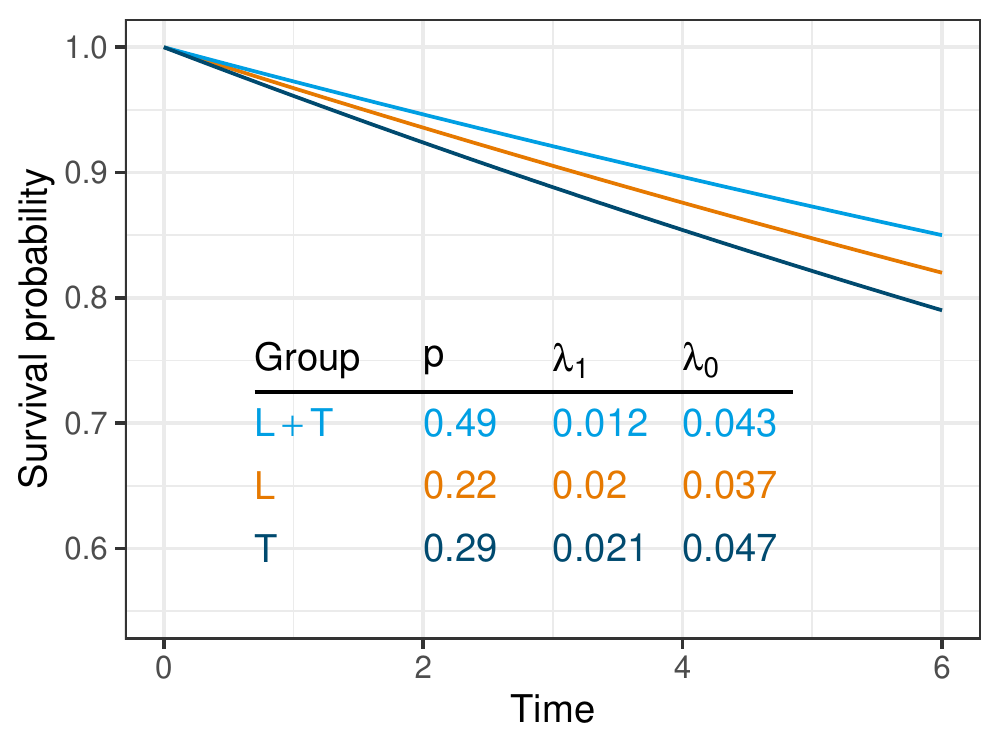} \caption{Distribution parameters and survival functions}\label{fig:ex-plot}
\end{figure}

Huober et al. (2019) investigated the effect of Lapatinib (L),
Trastuzumab (T), and a combination of both on pathologic complete
response and survival in patients with HER2-positive early breast
cancer. They reported group-wise response rates, and event-free as well
as overall survival rates. Furthermore, they estimated hazard ratios
between responders and non-responders within the treatment groups. Under
the assumption of exponentially distributed survival within the response
strata, the RSES distribution parameters can be derived. For overall
survival, they are given in Figure \ref{fig:ex-plot} together with the
survival functions. Under these assumptions, survival of responders is
considerably better in all groups. Due to the highest response
probability and the best survival for responders, the combination L+T
has the best overall survival. Even though a treatment with T leads to a
higher repsonse probability compared to L, the non-responder survival in
T is worse and the responder survival is almost equal. This results in a
better survival of L compared to T.

Table \ref{tab:ex-pow} shows approximate sample sizes per group for the
three pairwise comparisons between the three groups at global level 0.05
and global power 0.8. They are calculated for the exact test by the
method described in Section \ref{sec-ss-calc}. Furthermore, exact power
values for the exact test and simulation-based power values for the
logrank test and stratified logrank test are given. Compared to T, L+T
has a response advantage and a survival advantage in both strata and the
logrank test is more powerful than the exact test. In the other two
comparisons, the differences between response probability and stratified
survival are not uniform and partly compensate each other when compared
by the logrank test. Thus, the logrank test has lower power than the
exact test. Since the stratified logrank test deliberately ignores any
survival benefit arising from a response benefit, it has low power for
the comparison L+T vs.~T, although the survival curves show the greatest
difference. The power of the stratified logrank test for the comparison
between L + T and T is even lower due to the fact that it is not able to
capture the treatment effect mainly originating from the higher response
rate in L+T compared to L. However, when comparing T versus L, the
stratified logrank test is more powerful than the logrank test since it
does not struggle with the opposed effects on response and stratified
survival.

\begin{table}

\caption{\label{tab:ex-pow}Approximate sample sizes and power values of three group comparisons}
\centering
\begin{tabular}[t]{l|>{\raggedleft\arraybackslash}p{3cm}|>{\raggedleft\arraybackslash}p{3cm}|>{\raggedleft\arraybackslash}p{3cm}|>{\raggedleft\arraybackslash}p{3cm}}
\hline
Comparison & Approximate ample size & Power of exact test & Power of logrank test & Power of strat. logrank test\\
\hline
L+T vs. T & 86 & 0.79 & 0.87 & 0.37\\
\hline
L+T vs. L & 59 & 0.79 & 0.53 & 0.05\\
\hline
L vs. T & 378 & 0.80 & 0.33 & 0.76\\
\hline
\end{tabular}
\end{table}

\hypertarget{discussion}{%
\section{Discussion}\label{discussion}}

We derived some basic properties of the RSES model, constructed Maximum
Likelihood estimators, and developed an approximate and exact test. We
saw that the approximate test shows a Type I error inflation and almost
no power advantage compared to the exact test which makes the latter the
better choice. Xia, Cui, and Yang (2014) already showed that the logrank
test has low power if the treatment affects only response. Our work
showed that alternative methods like the RSES test can solve this
problem. Thus, survival analysis within the RSES model is not
recommended to be done with the commonly used logrank test. The
approximate sample size calculation method worked well and can be a good
start value for an easy to implement exact sample size calculation
method.

The main limitation of the derived methods is that they require the
absence of censoring. Integrating censored observations into the MLEs is
straightforward and doesn't affect the asymptotic distribution. Since
the presented confidence intervals and tests are based on the MLEs, they
can be applied without problems. However, exact calculations become much
more complicated and require specification of a censoring distribution.
We do not expect qualitatively different results regarding coverage
probabilities and error rates in the presence of censoring. The sample
size calculation method can be used to determine the needed number of
events which can then be corrected for censoring.

The strength of the fully parametric RSES model is that it allows the
exact derivation of statistical measures. For example, Xia, Cui, and
Yang (2014) calculated the correlation between the surrogate endpoint
and the overall survival. We calculated exact error probabilities for
our confidence intervals and test decisions.

The three-parameter-complexity of the RSES model can be seen as a
strength since it allows the consideration of complex treatment effects.
However, the unavoidable downside of this is that survival differences
might not be easily interpretable. This problem could be solved by using
a one-dimensional effect measure like the Restricted Mean Survival Time
(RMST). Also, estimation of stratum-specific survival parameters is
imprecise if the stratum is small. This is the case if the response
probability is near 0 or 1. In such cases, responder-stratified methods
should only be applied if the sample size is sufficiently large.

The fully parametric approach can also be seen as too restrictive.
Exponential distribution within the strata is a strong assumption that
is probably rarely met exactly. A less restrictive approach to
within-stratum estimation could use more flexible models like the
Weibull model or the non-parametric Kaplan-Meier estimator. By this,
however, interpretation of survival differences becomes even more
difficult.

In practice, response rates and stratum-specific Kaplan-Meier estimates
could be used to flexibly describe the treatment effect on response
probability and stratum survival. They can be aggregated at group-level
to descriptively compare survival of different treatment groups.
Confirmatory testing of survival difference can then be based on a
one-dimensional effect measure like RMST. For power analysis and sample
size calculation, stronger assumptions like in the RSES model can be
made. However, the circumstances are more complicated if the response
difference is assessed earlier in the study as a basis for an interim
decision on accelerated approval. Depending on the choice of global Type
I and Type II error control, the correlation of the test statistics of
response difference and survival difference has to be considered and
quantified in the planning stage. Our future research will focus on
analysis and sample size calculation in such study designs.

\hypertarget{references}{%
\section*{References}\label{references}}
\addcontentsline{toc}{section}{References}

\hypertarget{refs}{}
\begin{CSLReferences}{1}{0}
\leavevmode\vadjust pre{\hypertarget{ref-Boschloo1970}{}}%
Boschloo, R. D. 1970. {``Raised Conditional Level of Significance for
the 2 x 2-Table When Testing the Equality of Two Probabilities.''}
\emph{Statistica Neerlandica} 24: 1--9.
\url{https://doi.org/10.1111/j.1467-9574.1970.tb00104.x}.

\leavevmode\vadjust pre{\hypertarget{ref-Conforti2021}{}}%
Conforti, Fabio, Laura Pala, Isabella Sala, Chiara Oriecuia, Tommaso De
Pas, Claudia Specchia, Rossella Graffeo, et al. 2021. {``Evaluation of
Pathological Complete Response as Surrogate Endpoint in Neoadjuvant
Randomised Clinical Trials of Early Stage Breast Cancer: Systematic
Review and Meta-Analysis.''} \emph{BMJ} 375.
\url{https://doi.org/10.1136/bmj-2021-066381}.

\leavevmode\vadjust pre{\hypertarget{ref-FDA2020}{}}%
Food and Drug Administration. 2020. {``Guidance for Industry: Pathologic
Complete Response in Neoadjuvant Treatment of High-Risk Early-Stage
Breast Cancer: Use as an Endpoint to Support Accelerated Approval.''}
https://www.fda.gov/regulatory-information/search-fda-guidance-documents/pathological-complete-response-neoadjuvant-treatment-high-risk-early-stage-breast-cancer-use.

\leavevmode\vadjust pre{\hypertarget{ref-FDA2022}{}}%
---------. 2022. {``CDER Drug and Biologic Accelerated Approvals Based
on a Surrogate Endpoint as of December 31, 2021.''}
https://www.fda.gov/drugs/nda-and-bla-approvals/accelerated-approvals.

\leavevmode\vadjust pre{\hypertarget{ref-Huober2019}{}}%
Huober, Jens, Eileen Holmes, José Baselga, Evandro de Azambuja, Michael
Untch, Debora Fumagalli, Severine Sarp, et al. 2019. {``Survival
Outcomes of the NeoALTTO Study (BIG 1--06): Updated Results of a
Randomised Multicenter Phase III Neoadjuvant Clinical Trial in Patients
with Her2-Positive Primary Breast Cancer.''} \emph{European Journal of
Cancer} 118: 169--77.

\leavevmode\vadjust pre{\hypertarget{ref-Johnson1995}{}}%
Johnson, Norman L, Samuel Kotz, and Narayanaswamy Balakrishnan. 1995.
\emph{Continuous Univariate Distributions, Volume 2}. Vol. 289. John
Wiley \& Sons.

\leavevmode\vadjust pre{\hypertarget{ref-Mantel1966}{}}%
Mantel, Nathan. 1966. {``Evaluation of Survival Data and Two New Rank
Order Statistics Arising in Its Consideration.''} \emph{Cancer Chemother
Rep} 50: 163--70.

\leavevmode\vadjust pre{\hypertarget{ref-Mehrotra2003}{}}%
Mehrotra, Devan V., Ivan S. F. Chan, and Roger L. Berger. 2003. {``A
Cautionary Note on Exact Unconditional Inference for a Difference
Betwenn Two Independent Binomial Proportions.''} \emph{Biometrics} 59:
441--50. \url{https://doi.org/10.1111/1541-0420.00051}.

\leavevmode\vadjust pre{\hypertarget{ref-Wallach2018}{}}%
Wallach, Joshua D, Joseph S Ross, and Huseyin Naci. 2018. {``The US Food
and Drug Administration's Expedited Approval Programs: Evidentiary
Standards, Regulatory Trade-Offs, and Potential Improvements.''}
\emph{Clinical Trials} 15 (3): 219--29.

\leavevmode\vadjust pre{\hypertarget{ref-Xia2014}{}}%
Xia, Yi, Lu Cui, and Bo Yang. 2014. {``A Note on Breast Cancer Trials
with pCR-Based Accelerated Approval.''} \emph{Journal of
Biopharmaceutical Statistics} 24:5: 1102--14.
\url{https://doi.org/10.1080/10543406.2014.931410}.

\end{CSLReferences}

\appendix

\hypertarget{appendices}{%
\section{Appendices}\label{appendices}}

\hypertarget{surv-diff}{%
\subsection{Detailed analyses of survival differences}\label{surv-diff}}

Let \(p_i, \lambda_{1, i}, \lambda_{0, i}\) be the respective parameter
sets of the groups as shown in Figure \ref{fig:model}. Let \(S_i\) be
the survival functions of the two groups. We can differentiate three
cases of the relation of \(S_E\) and \(S_C\):

\begin{enumerate}
\def\labelenumi{\arabic{enumi}.}
\tightlist
\item
  Completely equal: \(S_E(t) = S_C(t) \quad \forall t \ge 0\)
\item
  Uniformly different: \(S_E(t) \ne S_C(t) \quad \forall t > 0\)
\item
  Crossing: not completely equal but \(\exists\ t > 0\) such that
  \(S_E(t) = S_C(t)\)
\end{enumerate}

It is easily seen that \(S_E\) and \(S_C\) are completely equal if and
only if one of the following conditions hold:

\begin{itemize}
\tightlist
\item
  \(p_E = p_C, \lambda_{1, E} = \lambda_{1, C}\) and
  \(\lambda_{0, E} = \lambda_{0, C}\)
\item
  \(p_E = p_C = 0\) and \(\lambda_{0, E} = \lambda_{0, C}\)
\item
  \(p_E = p_C = 1\) and \(\lambda_{1, E} = \lambda_{1, C}\)
\item
  \(\lambda_{1, E} = \lambda_{1, C} = \lambda_{0, E} = \lambda_{0, C}\)
\end{itemize}

These can be reformulated as:

\begin{itemize}
\tightlist
\item
  The parameter sets in both groups are equal.
\item
  There are no responders in both groups and the non-responder
  parameters are equal in both groups.
\item
  There are no non-responders in both groups and the responder
  parameters are equal in both groups.
\item
  Survival is equal for responder and non-responders and in both groups.
\end{itemize}

To investigate whether two survival curves cross for some \(t > 0\), we
have to compare the relation at \(t = 0\) with the relation at
\(t \to \infty\). The former is determined by the derivatives of \(S_E\)
and \(S_C\) at 0. The latter is determined by the minima of the hazards,
i.e.~the hazard of the fitter stratum (that is non-empty):
\[\lambda_{\min,G} := \min(\lambda_{1, G}, \lambda_{0, G}).\] In most
practical cases, this will be the responder stratum. In the special
cases \(p_G = 0 \ (1)\), set
\(\lambda_{\min,G} := \lambda_{0, G} \ (\lambda_{1, G})\). Define
\(\lambda_{\max,G}\) analogously. Let \(p_{G, \lambda_{\min}} = p_G\) if
\(\lambda_{\min,G} = \lambda_{1, G}\) and
\(p_{G, \lambda_{\min}} = 1-p_G\) if
\(\lambda_{\min,G} = \lambda_{0, G}\) be the proportion of the fitter
stratum. The curves don't cross if and only if one of the following
conditions is true (assuming the curves are not completely equal):

\begin{itemize}
\tightlist
\item
  The first non-equal derivatives at 0 fulfill
  \(S_E^{(k)}(0) > S_C^{(k)}(0)\) and one of the following statements is
  true:

  \begin{itemize}
  \tightlist
  \item
    \(\lambda_{\min,E} < \lambda_{\min,C}\)\\
  \item
    \(\lambda_{\min,E} = \lambda_{\min,C}\) and
    \(p_{E, \lambda_{\min}} > p_{C, \lambda_{\min}}\)\\
  \end{itemize}
\item
  Condition 2.1 with \(E\) and \(C\) exchanged.
\end{itemize}

This can be reformulated as:

\begin{itemize}
\tightlist
\item
  Event rate at the beginning is higher in group \(E\) and:

  \begin{itemize}
  \tightlist
  \item
    The fitter stratum in group \(E\) has better survival than the
    fitter stratum in group \(C\) or
  \item
    the fitter strata in both groups have equal survival but there are
    more responders in group \(E\) as compared to group \(C\).
  \end{itemize}
\item
  Condition 2.1 with \(E\) and \(C\) exchanged.
\end{itemize}

Two curves cross if and only if they are not completely equal and not
uniformly different.

\hypertarget{mle-derivation}{%
\subsection{Derivation of Maximum Likelihood
Estimators}\label{mle-derivation}}

The density function is
\[f_{p, \lambda_1, \lambda_0}(x, t) = x \cdot p \cdot \lambda_1\exp(-\lambda_1 t) + (1-x) \cdot (1-p) \cdot \lambda_0\exp(-\lambda_0 t).\]
Hence, the log likelihood of the three dimensional parameter
\((p, \theta_1, \theta_2)\) is:

\begin{align*}
\log \left(L_{\vec x, \vec t}(p, \theta_1, \theta_2)\right) 
& =  \sum\limits_{i = 1}^{k(\vec x)} \big[ \log(p) + \theta_1 - \exp(\theta_1)\cdot t_i \big] + \sum\limits_{i = k(\vec x)+1}^{n} \big[ \log(1-p) + \theta_0 - \exp(\theta_0)\cdot t_i \big] \\
& = k(\vec x)\cdot \log(p) + \left(n-k(\vec x)\right)\cdot\log(1-p) \\
& \quad\quad + \sum\limits_{i = 1}^{k(\vec x)} \big[ \theta_1 - \exp(\theta_1)\cdot t_i \big]
 + \sum\limits_{i = k(\vec x)+1}^{n} \big[ \theta_0 - \exp(\theta_0)\cdot t_i \big] 
\end{align*}

and can therefore be maximized within the summands. Finding the roots of
the derivatives of the summands yields the Maximum Likelihood estimators
shown in Section \ref{sec-mle}.

For variance approximation, we derive the values of the Fisher
information matrix by taking the negative expectation of the second
derivation of the log likelihood:
\[I(\vartheta)_{i, j} = -E\left[ \frac{d^2}{d\vartheta_i d\vartheta_j} \log L_{\vec X, \vec T}(\vartheta) \right]\]
We get \[I(\vartheta) = \begin{pmatrix}
-\frac{n}{p(1-p)} & 0 & 0 \\
0 & -n\cdot p & 0 \\
0 & 0 & -n\cdot (1-p)\\
\end{pmatrix}\] The inverse of this diagonal matrix is obtained by
taking the inverse of the diagonal entries. This yields the variance
estimators in Section \ref{sec-mle}.

\hypertarget{ci-cov-prob}{%
\subsection{Calculating exact coverage probability of confidence
intervals}\label{ci-cov-prob}}

The coverage probabilities of the asymptotic confidence intervals are
dependent on the true response probability \(p_0\). The coverage
probability of \(\mathrm{CI}_p\) can be calculated by
\[\mathrm{CP}(\mathrm{CI}_p) = \sum\limits_{k \in \{0, \dots, n\}} 1_{\left\{|p_0-k/n| \le z_{1-\alpha/2} \cdot \sqrt{k/n\cdot(1-k/n)/n}\right\}} \cdot f_{p_0}(k),\]
where \(f_{p_0}\) is the binomial density.

Since the distribution of \(\lambda_1\exp(-\hat \theta_1)\) conditional
on \(k\) is known to be \(\Gamma(k, k)\), we can calculate the coverage
probability of \(\mathrm{CI}_{\theta_1}\) conditional on \(k\) exactly
by
\[\mathrm{CP}_k(\mathrm{CI}_{\theta_1}) = F_{\Gamma(k, k)}\left(\exp\left(z_{1-\alpha/2}\sqrt{1/k}\right)\right) - F_{\Gamma(k, k)}\left(\exp\left(-z_{1-\alpha/2}\sqrt{1/k}\right)\right),\]
where \(F_{\Gamma(k, k)}\) is the distribution function of the
\(\Gamma(k, k)\)-distribution. The unconditional coverage probability
then is
\[\mathrm{CP}(\mathrm{CI}_{\theta_1}) = f_p(0) + \sum\limits_{k \in \{1, \dots, n\}} f_p(k)\cdot \mathrm{CP}_k(\mathrm{CI}_{\theta_1}).\]
The coverage probability of \(\mathrm{CI}_{\theta_0}\) can be calculated
analogously.

\hypertarget{exact-test-keeps-level}{%
\subsection{Exact test keeps Type I error
rate}\label{exact-test-keeps-level}}

Consider the test procedure described in Section \ref{sec-ex-test}. Then
the Type I error rate is:

\begin{align*}
P(\mathrm{Rej.}H_0) &= P(\mathrm{Rej.}H_{p, 0}) + P\left(\lnot\mathrm{Rej.}H_{p} \land (\mathrm{Rej.}H_{\theta_1, 0} \lor \mathrm{Rej.}H_{\theta_0, 0})\right)\\
&= P(\mathrm{Rej.}H_{p, 0}) + \sum\limits_{(k_E, k_C) \notin R_p} f(k_E, k_C) \cdot P(\mathrm{Rej.}H_{\theta_1, 0} \lor \mathrm{Rej.}H_{\theta_0, 0} | k_E, k_C) \\
&= P(\mathrm{Rej.}H_{p, 0}) + \sum\limits_{(k_E, k_C) \notin R_p} f(k_E, k_C) \cdot \left(1-P(\lnot\mathrm{Rej.}H_{\theta_1, 0} \land \lnot\mathrm{Rej.}H_{\theta_0, 0} | k_E, k_C)\right) \\
&= P(\mathrm{Rej.}H_{p, 0}) + \sum\limits_{(k_E, k_C) \notin R_p} f(k_E, k_C) \cdot \left(1-P(\lnot\mathrm{Rej.}H_{\theta_1, 0}| k_E, k_C) \cdot P(\lnot\mathrm{Rej.}H_{\theta_0, 0} | k_E, k_C)\right) \\
&\le P(\mathrm{Rej.}H_{p, 0}) + \sum\limits_{(k_E, k_C) \notin R_p} f(k_E, k_C) \cdot \left(1- (1-\tilde\alpha) \cdot (1-\tilde\alpha)\right) \\
&= P(\mathrm{Rej.}H_{p, 0}) + \left(1 - P(\mathrm{Rej.}H_{p, 0})\right) \cdot (1- (1-\tilde\alpha)^2) \\
&= 1 - \left(1 - P(\mathrm{Rej.}H_{p, 0})\right) \cdot (1-\tilde\alpha)^2 \\
&\le 1 - (1-\tilde\alpha)^3 \\
&= \alpha
\end{align*}

The equality in the fourth line holds because conditional on \(k_E\) and
\(k_C\), \(T_{\theta_1}\) and \(T_{\theta_0}\) are independent. The
inequality in the fifth line is actually an equality because
\(T_{\theta_1}\) and \(T_{\theta_0}\) have a continuous distribution and
hence their exact tests exploit the local level.

\hypertarget{logrank-tests}{%
\subsection{Logrank test statistics}\label{logrank-tests}}

We use the logrank test statistic introduced by (Mantel 1966) with
hypergeometric variance estimation. Let \(t^{(i)}\) denote the event
times. Let \(Y^{(i)}\) be the total number at risk and \(Y_E^{(i)}\) the
number at risk in the experimental group immediately before \(t^{(i)}\).
Let \(d^{(i)}\) be the total number of events and \(d_E^{(i)}\) the
number of events in the experimental group at \(t^{(i)}\). Then
\[E^{(i)} = d \cdot \frac{Y_E^{(i)}}{Y^{(i)}}\] is the expected number
of events in the experimental group at \(t^{(i)}\). The conditional
variance of \(d_E^{(i)}\) is derived from the hypergeometric
distribution as
\[V^{(i)} = \frac{(Y^{(i)} - Y_E^{(i)}) \cdot Y_E^{(i)} \cdot (Y^{(i)} - d^{(i)})\cdot d^{(i)}}{{Y^{(i)}}^2 \cdot (Y^{(i)} - 1)}.\]
The total number of observed and expected events are
\[O = \sum_i d_E^{(i)}\] and \[E = \sum_i E^{(i)}.\] The approximate
variance of \(O-E\) is \[V = \sum_i V^{(i)}.\] The logrank test
statistic then is \[T_{\mbox{LR}} = \frac{O-E}{\sqrt{V}}.\]

For the stratified logrank test, the quantities \(O_j, E_j\) and \(V_j\)
are calculated within each stratum \(j\). The test statistic then is
\[T_{\mbox{sLR}} = \frac{\sum_j (O_j - E_j)}{\sqrt{\sum_j V_j}}.\]

\hypertarget{appr-acc-prob}{%
\subsection{Derivation of approximate acceptance probabilities of local
hypotheses}\label{appr-acc-prob}}

Approximate acceptance probabilities of the three local tests can be
derived using the asymptotic normality of the test statistics. Let
\(n_E, n_C\) be the sample sizes and
\(p_i^{\prime}, \theta_{j, i}^{\prime}\) the specified parameter values
under the assumed alternative hypothesis. Then,
\[\hat p \approx \tilde p := \frac{n_E p_E^{\prime}+ n_C p_C^{\prime}}{n_E + n_C}\]
and
\[ \mathrm{Var}(\hat p_E - \hat p_C) = {\sigma_p^{\prime}}^2 := \frac{p_E^{\prime}(1-p_E^{\prime})}{n_E} + \frac{p_C^{\prime}(1-p_C^{\prime})}{n_C}. \]
Hence, under the alternative we have
\[ T_p \overset{\mathrm{appr}}{\sim}N\left(\frac{p_E^{\prime}-p_C^{\prime}}{\sqrt{\tilde p(1- \tilde p)(\frac{1}{n_E} + \frac{1}{n_C})}}, \frac{{\sigma_p^{\prime}}^2}{\tilde p(1- \tilde p)(\frac{1}{n_E} + \frac{1}{n_C})}\right). \]
Thus, for \(p_E^{\prime}\ge p_C^{\prime}\), the acceptance probability
for a two-sided test of \(H_{p, 0}\) at level \(\alpha\) is
approximately
\[\beta_p = \Phi\left( \frac{z_{1-\frac{\alpha}{2}} \cdot \sqrt{\tilde p(1- \tilde p)(\frac{1}{n_E} + \frac{1}{n_C})} - (p_E^{\prime}-p_C^{\prime})}{\sigma_p^{\prime}}\right).\]

The variance of the numerator of \(T_{\theta_1}\) is approximately
\[ \mathrm{Var}(\hat\theta_{1, E} - \hat\theta_{1, C}) \approx {\sigma_{\theta_1}^{\prime}}^2 := \frac{1}{n_E p_E^{\prime}} + \frac{1}{n_C p_C^{\prime}}. \]
Hence, under the alternative we have
\[ T_{\theta_1} \overset{\mathrm{appr}}{\sim}N\left(\frac{\theta_{1, E}^{\prime}- \theta_{1, C}^{\prime}}{\sqrt{\tfrac{1}{\tilde p}\cdot(\tfrac{1}{n_E} + \tfrac{1}{n_C})}}, \frac{{\sigma_{\theta_1}^{\prime}}^2}{\tfrac{1}{\tilde p}\cdot(\tfrac{1}{n_E} + \tfrac{1}{n_C})} \right). \]
Thus, for \(\theta_{1, E}^{\prime}\le \theta_{1, C}^{\prime}\) the power
for a two-sided test at level \(\alpha\) is approximately
\[\beta_{\theta_1} = \Phi\left( \frac{z_{1-\tfrac{\alpha}{2}} \cdot \sqrt{\tfrac{1}{\tilde p}\cdot(\tfrac{1}{n_E} + \tfrac{1}{n_C})} - (\theta_{1, C}^{\prime}-\theta_{1, E}^{\prime})}{\sigma_{\theta_1}^{\prime}}\right)\]

Analogously, for \(\theta_{0, E}^{\prime}\le \theta_{0, C}^{\prime}\),
the acceptance probability for a two-sided test of \(H_{\theta_0, 0}\)
at level \(\alpha\) is approximately
\[\beta_{\theta_0} = \Phi\left( \frac{z_{1-\tfrac{\alpha}{2}} \cdot \sqrt{\tfrac{1}{(1-\tilde p)}\cdot(\tfrac{1}{n_E} + \tfrac{1}{n_C})} - (\theta_{0, C}^{\prime}-\theta_{0, E}^{\prime})}{\sigma_{\theta_0}^{\prime}}\right).\]

\bibliographystyle{unsrt}
\bibliography{bibliography.bib}

\end{document}